\documentclass[a4paper,fleqn,usenatbib]{mnras}

\usepackage{newtxtext,newtxmath}

\usepackage[T1]{fontenc}
\usepackage{ae,aecompl}

\usepackage[dvipdfmx]{graphicx}	
\usepackage{xspace}
\usepackage{booktabs,caption}
\usepackage{threeparttable}

\usepackage{color}
\usepackage{ulem}
\newcommand{\fdir}{./}

\newcommand{\msun}{M_\odot}

\newcommand{\rsun}{R_\odot}
\newcommand{\zsun}{Z_\odot}
\newcommand{\mmax}{M_{\max}}

\newcommand{\cago}{^{12}{\rm C}(\alpha,\gamma)^{16}{\rm O}}
\newcommand{\tsig}{3\sigma}
\newcommand{\euclid}{{\it Euclid}\xspace}

\title[\euclid detectability]{\euclid detectability of pair
  instability supernovae in binary population synthesis models
  consistent with merging binary black holes}

\author[Ataru Tanikawa]{ Ataru Tanikawa$^{1,}$\thanks{E-mail:
    tanikawa@ea.c.u-tokyo.ac.jp}, Takashi J. Moriya$^{2,3,4}$, Nozomu
  Tominaga$^{2,3,5,6}$, Naoki Yoshida$^{7,6,8}$ \\
$^{1}$Department of Earth Science and Astronomy, College of
  Arts and Sciences, The University of Tokyo, 3-8-1 Komaba, Meguro-ku,
  Tokyo 153-8902, Japan\\
$^{2}$National Astronomical Observatory of Japan, National
  Institutes of Natural Sciences, 2-21-1 Osawa, Mitaka, Tokyo
  181-8588, Japan\\
$^{3}$Department of Astronomical Science, School of Physical Sciences,
  The Graduate University of Advanced Studies (SOKENDAI), 2-21-1
  Osawa, Mitaka, \\Tokyo 181-8588, Japan\\
$^{4}$School of Physics and Astronomy, Faculty of Science,
  Monash University, Clayton, Victoria 3800, Australia\\
$^{5}$Department of Physics, Faculty of Science and
Engineering, Konan University, 8-9-1 Okamoto,
Kobe, Hyogo 658-8501, Japan\\
$^{6}$Kavli Institute for the Physics and Mathematics of the Universe
  (WPI), The University of Tokyo, 5-1-5 Kashiwanoha, Kashiwa, Chiba
  277-8583, Japan\\
$^{7}$Department of Physics, School of Science, The University
  of Tokyo, 7-3-1 Hongo, Bunkyo, Tokyo 113-0033, Japan\\
$^{8}$Research Center for the Early Universe, School of
  Science, The University of Tokyo, 7-3-1 Hongo, Bunkyo, Tokyo
  113-0033, Japan}

\date{Accepted XXX. Received YYY; in original form ZZZ}

\pubyear{2017}

\begin{document}
\label{firstpage}
\pagerange{\pageref{firstpage}--\pageref{lastpage}}
\maketitle

\begin{abstract}

  We infer the expected detection number of pair instability
  supernovae (PISNe) during the operation of the \euclid space
  telescope based on binary population models. Our models reproduce
  the global maximum of the rate at the primary BH mass of $\sim 9-10$
  $\msun$, and the overall gradient of the primary BH mass
  distribution in the binary BH merger rate consistent with recent
  observations. We consider different PISN conditions depending on the
  $\cago$ reaction rate. The fiducial and $\tsig$ models adopt the
  standard and $\tsig$-smaller reaction rate, respectively. Our
  fiducial model predicts that \euclid detects several hydrogen-poor
  PISNe.  For the $\tsig$ model, detection of $\sim 1$ hydrogen-poor
  PISN by \euclid is expected if the stellar mass distribution extends
  to $\mmax = 600 \msun$, but the expected number becomes
  significantly smaller if $\mmax = 300 \msun$. We may be able to
  distinguish the fiducial and $\tsig$ models by the observed PISN
  rate. This will help us to constrain the origin of binary BHs and
  the reaction rate, although there remains degeneracy between $\mmax$
  and the reaction rate. PISN ejecta mass estimates from light curves
  and spectra obtained by follow-up observations would be important to
  disentangle the degeneracy.

\end{abstract}


\section{Introduction}
\label{sec:Introduction}

Pair instability supernovae (PISNe) are theoretically predicted as
thermonuclear explosions of very massive stars, leaving behind no
stellar remnants \citep[e.g.][]{1964ApJS....9..201F,
  1967PhRvL..18..379B, 1968Ap&SS...2...96F, 1983A&A...119...61O,
  1984ApJ...280..825B, 1986A&A...167..274E, 2001ApJ...550..372F,
  2002ApJ...567..532H, 2002ApJ...565..385U, 2012ARA&A..50..107L}.
PISNe have not been conclusively discovered despite of their
importance, although there are some possible candidate
\citep[e.g.,][]{2017NatAs...1..713T}. Near-solar metallicity stars are
hard to keep such massive helium cores until their deaths
\citep{Yoshida14} unless highly magnetized
\citep{2017A&A...599L...5G}, and only low-metallicity stars are
suggested to explode as PISNe \citep{2007A&A...475L..19L}. Since
low-metallicity stars are mainly formed at high redshift,
near-infrared (NIR) observatories should be appropriate for PISN
surveys. \cite{2012MNRAS.422.2701P}, \cite{2019PASJ...71...59M,
  2022ApJ...925..211M, 2022A&A...666A.157M},
\cite{2019PASJ...71...60W} and \cite{2020ApJ...894...94R} have
predicted that PISNe can be discovered by NIR observatories:
ULTIMATE-Subaru\footnote{\url{https://www.naoj.org/Projects/newdev/ngao/}},
the Nancy Grace Roman Space Telescope \citep{2015arXiv150303757S}, the
\euclid space telescope \citep{2011arXiv1110.3193L}, and the James
Webb Space Telescope \citep{2006SSRv..123..485G, 2013ApJ...777..110W}.

PISNe can play an important role in shaping black hole (BH) mass
distribution in the universe and in binary BHs observed by
gravitational wave (GW) observatories
\citep[e.g.][]{2021arXiv211103606T}. They are predicted to form a mass
gap in BH mass distribution \citep{2016A&A...594A..97B,
  2017ApJ...836..244W, 2017MNRAS.470.4739S, 2018MNRAS.474.2959G,
  2019ApJ...882...36M, 2020ApJ...897..100V, 2022MNRAS.516.2252O},
hereafter the PI mass gap. The PI mass gap range has been under
debate, since it depends on the $\cago$ reaction rate
\citep{2018ApJ...863..153T, 2020ApJ...902L..36F, 2021MNRAS.501.4514C,
  2021ApJ...912L..31W, 2022ApJ...924...39M}, stellar metallicity
\citep{2019ApJ...887...53F, 2021MNRAS.501L..49K, 2021MNRAS.502L..40F,
  2021MNRAS.505.2170T}, stellar wind strength
\citep{2020ApJ...890..113B, 2021MNRAS.504..146V}, and stellar rotation
effect \citep{1985A&A...149..413G, 2020ApJ...888...76M,
  2020A&A...640L..18M}. The PI mass gap can be buried to some degree
if binary BHs are formed through dynamical capture
\citep{2019PhRvD.100d3027R, 2020MNRAS.497.1043D, 2020ApJ...903L..40L,
  2020ApJ...904L..13R, 2021ApJ...908..194T, 2021MNRAS.501.5257R,
  2022MNRAS.516.1072C, 2022arXiv220403493B}. There are many other
suggestions for filling the PI mass gap \citep{2020arXiv200707889C,
  2020ApJ...903L..21S, 2021PhRvD.104d3015Z,
  2021arXiv211103094S}. Nevertheless, PISNe can have great impacts on
binary BH population.

\cite{2022A&A...666A.157M} recently investigated the expected numbers
of PISN discovery with the \euclid space telescope. Their discovery
number estimates are based on the PISN event rates that are scaled
with the observed event rates of superluminous SNe
\citep{2013MNRAS.431..912Q, 2017MNRAS.464.3568P}. In this paper, we
predict the expected PISN detection number by \euclid, based on two
sets of binary population synthesis models consistent with binary BHs
observed by GWs with respect to the global maximum of the rate at the
primary BH mass of $\sim 9-10$ $\msun$, and the global gradient of the
primary BH mass distribution in the binary BH merger rate.
\citep{2022ApJ...926...83T}. The two models adopt different PISN
models in which PISNe happen in stars with helium core mass of
$65-135$ and $90-180 \msun$. The former and latter models assume the
$\cago$ reaction rate as the median value of the STARLIB
\citep{2013ApJS..207...18S}, and one-third of that (or smaller than
that by $\tsig$), respectively. Since PISNe are caused by stars with
different masses in the two model series, we obtain different PISN
detection numbers. We show that the \euclid PISN survey will be
helpful to constrain the formation mechanism of binary BHs among a
considerable number of suggested scenarios
\citep{2021arXiv211103634T}, and the $\cago$ reaction rate. These
models do not fully reproduce the details of the primary BH mass
distribution in binary BH mergers, such as the second peak of the
binary BH merger rate at the primary BH mass of $\sim 35$ $\msun$
\citep{2021arXiv211103634T}. Nevertheless, it is worth while to
investigate the models, since they can reproduce the PI mass gap event
GW190521 \citep{2020PhRvL.125j1102A, 2020ApJ...900L..13A}, despite
considering only the PI mass gap event formed from isolated binary
stars.

\section{Method}
\label{sec:Method}

We exploit the fiducial and L-$\tsig$ models in
\cite{2022ApJ...926...83T}. We first explain these previous
models. For both models we adopt the models of
\cite{2017ApJ...840...39M} (MF17) and \cite{2020MNRAS.492.4386S}
(SW20) for Population (Pop) I/II and Pop III star formation histories
(SFHs), respectively. The average metallicity of Pop I/II evolves in
the same way as MF17, and the metallicity is distributed as the
log-normal distribution centered on the average metallicity and
dispersion of $0.35$ at each time. The initial stellar mass functions
(IMFs) of single stars and primary stars of binary stars gradually
transition from a top-light IMF to a top-heavy IMF with metallicity
decreasing, which is motivated by numerical results of
\cite{2021MNRAS.508.4175C}. The maximum stellar mass ($\mmax$) is $150
\msun$ for all the metallicities. We perform binary population
synthesis simulation by means of {\tt
  BSEEMP}\footnote{\url{https://github.com/atrtnkw/bseemp}}
\citep{2020MNRAS.495.4170T, 2021MNRAS.505.2170T}, which is based on
the {\tt BSE} code developed by \cite{2000MNRAS.315..543H,
  2002MNRAS.329..897H}. Different points between the fiducial and
L-$\tsig$ models are as follows. The fiducial and L-$\tsig$ models
adopt the M and L models, respectively, for single star evolution with
metallicity $Z/\zsun \le 0.1$ ($\zsun = 0.02$), while both models
adopt the {\tt BSE} original models for single star evolution with
$Z/\zsun > 0.1$. Their features can be seen in appendix of
\cite{2022ApJ...926...83T}. In the fiducial and L-$\tsig$ models,
PISNe happen for helium core mass of $65-135 \msun$ and $90-180
\msun$, respectively. Note that we also consider pulsational PISNe
\citep[PPISNe][]{2002ApJ...567..532H, 2007Natur.450..390W,
  2012ARA&A..50..107L, 2014ApJ...792...28C, 2016MNRAS.457..351Y,
  2017ApJ...836..244W, 2019ApJ...882...36M, 2019ApJ...887...53F,
  2019ApJ...887...72L, 2020A&A...640A..56R}, which occur for helium
core mass of $45-65 \msun$, and leave behind $45$ $\msun$ BHs in the
fiducial model. On the other hand, they do not occur in the L-$\tsig$
model.  The different mass ranges are motivated by effects of the
$\cago$ nuclear reaction rate on star evolution
\citep{2018ApJ...863..153T,2020ApJ...902L..36F}.

\begin{table}
  \centering
  \begin{threeparttable}
    \caption{Summary of the previous and current
      models.} \label{tab:Model}
    \begin{tabular}{l|ccccc}
      \hline
      Name & Star & PPISN & PISN & SFH & $\mmax$ \\
      \hline
      \multicolumn{2}{l}{Previous models} \\
      fiducial     & M & $45-65$ & $65-135$ & MF17+SW20 & $150$ \\
      L-$\tsig$    & L & N/A     & $90-180$ & MF17+SW20 & $150$ \\
      \hline
      \multicolumn{2}{l}{Current models} \\
      fid.150      & M & $45-65$ & $65-135$ & H22+SW20  & $150$ \\
      fid.300      & M & $45-65$ & $65-135$ & H22+SW20  & $300$ \\
      $\tsig$.150  & L & N/A     & $90-180$ & H22+SW20  & $150$ \\
      $\tsig$.300  & L & N/A     & $90-180$ & H22+SW20  & $300$ \\
      $\tsig$.600  & L & N/A     & $90-180$ & H22+SW20  & $600$ \\
      \hline
    \end{tabular}
    \begin{tablenotes}
      \small
    \item The ``star'' column indicate adopted single star models
      for $Z/\zsun \le 0.1$. The ``PPISN'', and ``PISN'' columns
      indicate the mass ranges of helium cores generating PPISNe and
      PISNe, respectively. The units of the ``PPISN'', ``PISN''
      and ``$\mmax$'' columns are $\msun$. The fiducial and L-$\tsig$
      models (previous models) are used in
      \cite{2022ApJ...926...83T}. Other 5 models (current models) are
      used in this paper. Except for the Pop I/II SFH, the fid.150 and
      $\tsig$.150 models are identical to the fiducial and L-$\tsig$
      models, respectively.
    \end{tablenotes}
  \end{threeparttable}
\end{table}

We modify these models as follows. We choose the model of
\cite{2022ApJS..259...20H} (H22) for Pop I/II SFH instead of the
MF17's model. The H22's model is based on higher-redshift ($z$)
observations than the MF17's model, and consistent with the results of
\cite{2018ApJ...855..105O}. The star formation rate density of the
H22's model is smaller than the MF17's model at $z \gtrsim 8$. We
prepare initial conditions with larger $\mmax$ besides the original
ones with $\mmax = 150 \msun$ in order to make clear the dependence of
PISN detection rates on $\mmax$. We summarize the previous and current
models in Table \ref{tab:Model}. The current model series ``fid'' and
``$\tsig$'' correspond to the previous fiducial and L-$\tsig$ models,
respectively.

We briefly present parameters related to stable mass transfer and
common envelope evolution, which play crucial roles for forming binary
BHs. We choose these parameters in the same way as in
\cite{2022ApJ...926...83T}. The stability for stable mass transfer and
common envelope evolution is determined by the mass ratio of binary
stars depending on whether a donor star has radiative or convective
envelope. The mass ratio is given as in {\tt BSE}. The criteria for a
stellar envelope is different between {\tt BSE} and {\tt BSEEMP}. Core
(shell) helium burning stars have radiative (convective) envelopes in
{\tt BSE}, while stars with $\log T_{\rm eff}/{\rm T} \ge 3.65$
($<3.65$) have radiative (convective) envelopes in {\tt BSEEMP}, where
$T_{\rm eff}$ is effective temperature. The formulae of stable mass
transfer are the same as in {\tt BSE}. In stable mass transfer, the
maximum fraction of transferred mass is 0.5. We adopt the $\alpha$
formalism for the common envelope evolution
\citep{1984ApJ...277..355W}, and set $\alpha=1$ and $\lambda$ of
\cite{2014A&A...563A..83C}, where $\lambda$ is a numerical factor of
the binding energy of a stellar envelope, and depends on the envelope
structure \citep{1990ApJ...358..189D}. If a donor star is in the
Hertzsprung gap phase, and its mass transfer is unstable, we assume
that it merges with its companion star. This is because such a star
does not have steep density gradient between its core and envelope
\citep{2004ApJ...601.1058I}.

Because we focus on PISNe in this {\it Letter}, we describe our model
relevant to PISNe in more detail. We construct single star evolution
models up to $1280 \msun$, based on simulation results by the {\tt
  HOSHI} code \citep{2016MNRAS.456.1320T, 2018ApJ...857..111T,
  Takahashi19, Yoshida19}. The {\tt HOSHI} code can generate 2 types
of single star models referred as the M and L models, which are
similar to the GENEC without rotation \citep{Ekstroem12} and Stern
\citep{2011A&A...530A.115B}, respectively. Details of generation of
the M and L models are described in Appendix A of
\cite{2022ApJ...926...83T}. Here, we describe the treatment of
convective overshoot, which is important for the difference between
the M and L models. Actually, the difference between the M and L
models is only in the convective overshoot parameter; the parameter
for the M model is smaller than for the L model. This makes the
maximum radii of stars different in some stellar mass and metallicity
ranges. For example, Pop III stars with $90$ $\msun$ expand to $40$
and $3000$ $\rsun$ in the M and L models, respectively. We treat the
convective overshoot as a diffusive process above convective
regions. The diffusion coefficient of the convective overshoot
exponentially decreases with the distance from the convective boundary
as
  \begin{align}
    D_{\rm cv}^{\rm ov} = D_{\rm cv,0} \exp \left(-2 \frac{\Delta
      r}{f_{\rm ov} H_{\rm P0}} \right),
  \end{align}
  \citep{2000A&A...360..952H} where $D_{\rm cv,0}$ and $H_{\rm p0}$
  are the diffusion coefficient and the pressure scale height at the
  convective boundary, respectively, and $\Delta r$ is the distance
  from the convective boundary. The overshoot parameter $f_{\rm ov}$
  is set to be $0.01$ and $0.03$ for the M and L models
  \citep{Yoshida19}. As seen above, the M model considers inefficient
  (not without) convective overshoot, because ab initio hydrodynamical
  simulations show convective boundary mixing
  \citep{2022ApJ...926..169A}. We stop our single star evolution
  calculations at the carbon ignition time, although the {\tt HOSHI}
  code can follow later star evolution, such as carbon burning and PI
  mass loss.  We assume that supernova mass loss (including PPISN and
  PISN mass loss) occurs instantaneously at the stopping time. We
  adopt this modeling in order to match with those in the {\tt BSE}
  original models. Single star evolution after the carbon ignition
  might affect binary star evolution in reality.

We take into account stellar winds modeled by
\cite{2010ApJ...714.1217B} with mass loss of luminous stars
\citep{Nieuwenhuijzen90, 1989A&A...219..205K}, the NJ wind for
short. The NJ wind is used when stars are both blue and red
supergiants. Under this stellar wind model, a star with $\sim 300
\msun$ and $\sim 0.01 \zsun$ experiences mass loss of $\sim
10^{-3}~\msun~{\rm yr}^{-1}$ at its red supergiant phase due to the NJ
wind. This mass loss rate is similar to the pulsation-driven mass loss
\citep[see fig. 5 in][]{2020ApJ...902...81N}. This NJ wind mass loss
strips a hydrogen envelope (i.e. about half mass of a zero-age
main-sequence star) from a $\gtrsim 100 \msun$ star with $0.1 \zsun$
and a $\gtrsim 200 \msun$ star with $0.01 \zsun$. We adopt the Fryer's
rapid model \citep{2012ApJ...749...91F} with PPISN and PISN
effects. Note that the Fryer's rapid model does not form BHs with
$\lesssim 6 \msun$ unlike the Fryer's delayed model
\citep{2012ApJ...749...91F} or their latest model
\citep{2022ApJ...931...94F}. The PPISN and PISN effects are similar to
those in \cite{2016A&A...594A..97B} for the fiducial model, and to
those in \cite{2020ApJ...905L..15B} or ``the revised PSN'' model in
\cite{2022MNRAS.516.2252O} for the $\tsig$ model.

We gather PISN events from both single and binary stars. The numbers
of single and binary stars are equal, and thus the intrinsic binary
fractions in our models are 1, similar to an intrinsic binary fraction
$0.69^{+0.09}_{-0.09}$ derived by \cite{2012Sci...337..444S}. We take
into account PISNe from close and interacting binary stars, wide and
non-interacting binary stars, disrupted binary stars, and binary
merger products. Case A mergers yield more massive main-sequence stars
than their pre-merger stars, which can form more massive helium cores
than their pre-merger stars at their post main-sequence stage. This
can increase PISNe \citep[e.g.][]{2019ApJ...876L..29V}. On the other
hand, case B merger products have the same helium core mass as the
pre-merger stars. They may avoid PISNe if their helium core masses are
small \citep{2020MNRAS.497.1043D, 2020ApJ...904L..13R,
  2022MNRAS.516.1072C, 2022arXiv220403493B}. Mass transfer and mergers
rejuvenate the accretors and merger products, respectively, in the
same way as in {\tt BSE} as follows. Regardless of case A and B mass
transfers, the age of an accretor is calculated so as to keep the
fraction of main-sequence lifetime constant before and after the
accretor increases its mass \citep[see the details in section 7.1
  of][]{2000MNRAS.315..543H}\footnote{\cite{2000MNRAS.315..543H}
describe how to determine an age of a star decreasing its mass through
stellar wind in section 7.1. However, this also applies to how to
determine an age of a star increasing its mass through mass
transfer.}. The age of a case A merger product is determined as
eq. (80) of \cite{2002MNRAS.329..897H}. The age of a case B merger
product is the same as the age of the pre-merger post main-sequence
star. These modelings produce the following results in our
simulations. Through case A or B mass transfers, a star forms a more
massive helium core than a star not undergoing these mass
transfers. Similarly, case A merger products form more massive helium
cores than the pre-merger stars. This is because of the mass increase,
not because of the rejuvenation. Rejuvenation itself is not effective
on whether the number of PISNe increases or decreases. In {\tt BSE},
helium core masses depend only on the terminal-age main-sequence
masses, not on stellar evolution histories like
rejuvenations\footnote{In reality, stellar evolution histories can
change helium core masses. This is because stellar internal structure
can be affected by merger and mass accretion
\citep{1983Ap&SS..96...37H, 1984Ap&SS.104...83H,
  2021ApJ...923..277R}.}. For a case B merger, the merger product is
not rejuvenated.

\begin{figure}
  \includegraphics[width=\columnwidth]{\fdir/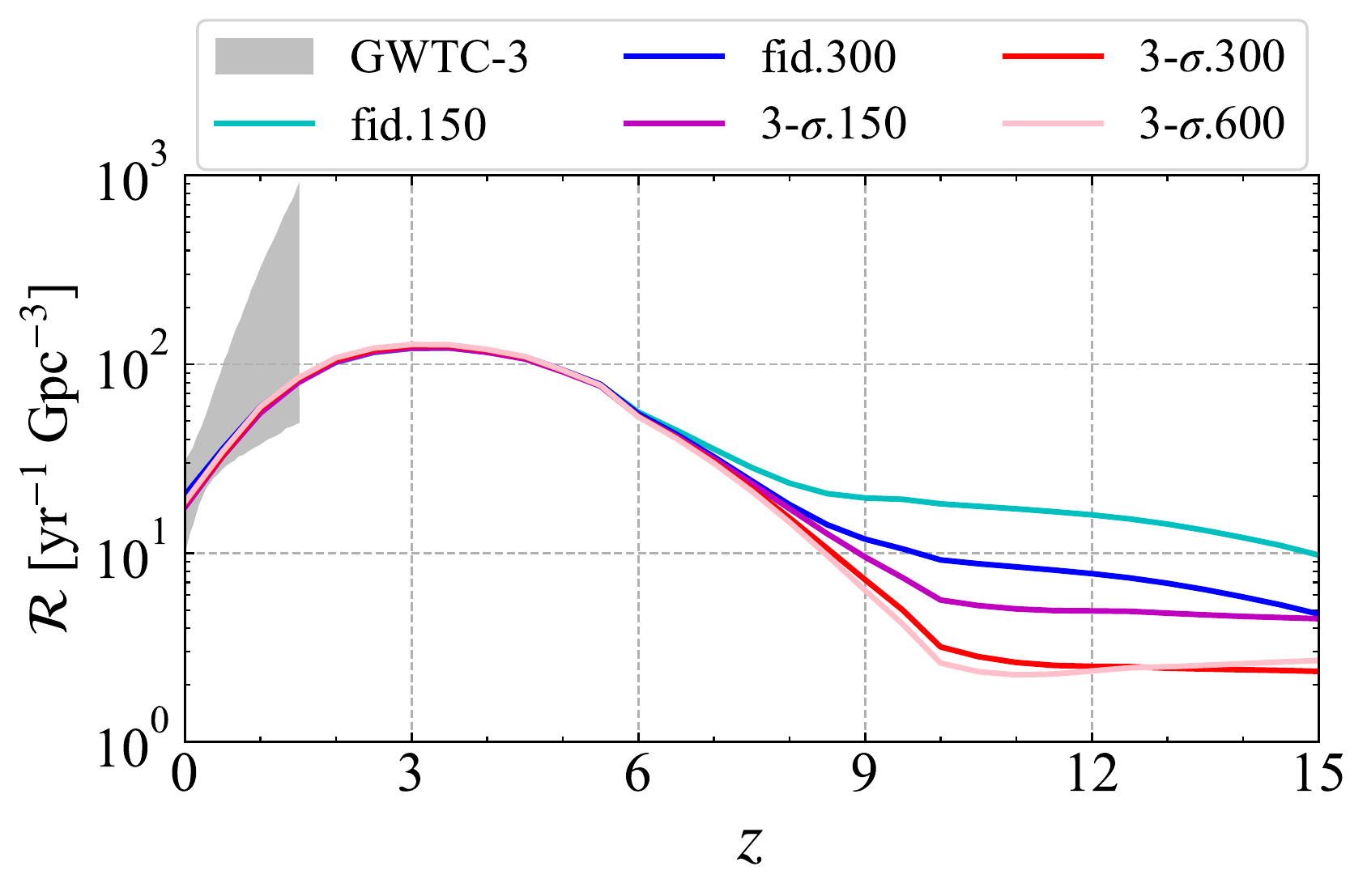}
  \caption{Redshift evolution of binary BH merger rate density (${\cal
      R}$) for the current 5 models shown in Table
    \ref{tab:Model}. The gray-shaded regions shows the 90 \% credible
    interval inferred by GWTC-3.}
  \label{fig:mergerRateFromFileTotal}
\end{figure}

\begin{figure}
  \includegraphics[width=\columnwidth]{\fdir/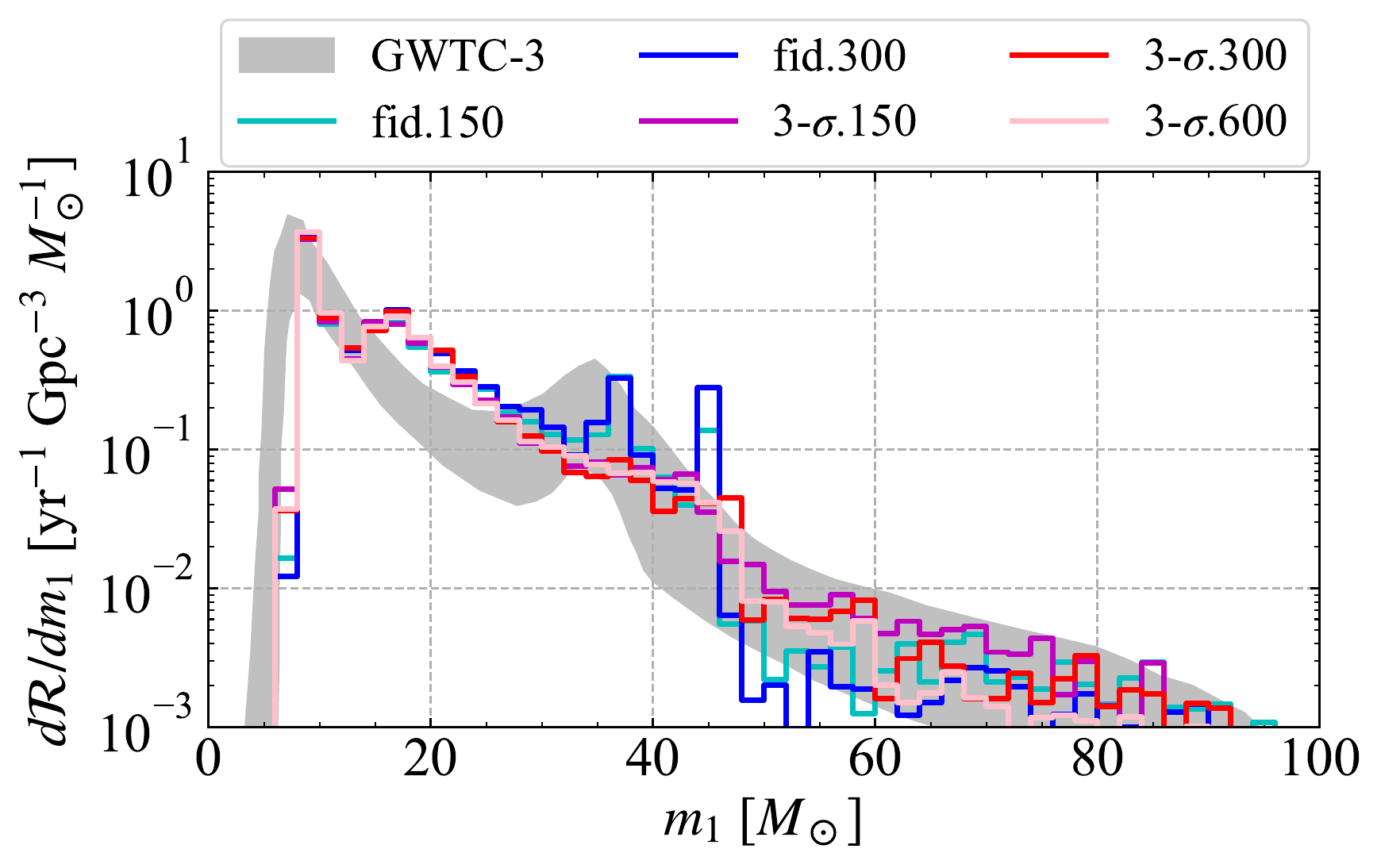}
  \caption{Primary BH mass distribution of merging binary BHs at $z=0$
    ($d{\cal R}/dm_1$) for the current 5 models shown in Table
    \ref{tab:Model}. The gray-shaded regions shows the 90 \% credible
    interval inferred by GWTC-3.}
  \label{fig:mergerRateFromFileMass}
\end{figure}

Figures \ref{fig:mergerRateFromFileTotal} and
\ref{fig:mergerRateFromFileMass} show the redshift evolution of binary
BH merger rate density (${\cal R}$) and primary BH mass distribution
of merging binary BHs at $z=0$ ($d{\cal R}/dm_1$), respectively, for
the current 5 models. In all the current models, the redshift
evolution and primary BH mass distribution matches well with the 90 \%
credible interval inferred by the LIGO-Virgo Gravitational-Wave
Transient Catalog 3 (GWTC-3), despite that these models are slightly
different from the previous models in
\cite{2022ApJ...926...83T}. Taking a look at the primary BH mass
distribution, we can see that the current 5 models reproduce the
global maximum at $\sim 9-10$ $\msun$, and the global gradient of the
primary BH mass distribution in the binary BH merger rate from $\sim
10 \msun$ to $\sim 100 \msun$. Notably, all the current models form
binary BH mergers with $65-100 \msun$ BHs, the so-called PI mass gap
BHs, like GW190521 \citep{2020ApJ...900L..13A,
  2020PhRvL.125j1102A}. Such PI mass BHs are formed from Pop III
binary stars in model series ``fid'' \citep{2021MNRAS.505.2170T},
while their origins can be Pop II binary stars in model series
``$\tsig$'' owing to upward shift of the lower mass bound of PISNe
\citep[e.g.][]{2020ApJ...905L..15B}.

We recognize that the current 5 models are not always consistent with
the bump at $\sim 35 \msun$ \citep[see the detail discussions
  in][]{2022ApJ...931...17V, 2022RNAAS...6...25R}. Although the
fiducial models show such bumps, they may disappear depending on the
model details.  We find that around the bump merging BHs are equally
formed through stable mass transfer channel and through common
envelope channel due to a small $\lambda$ parameter of common
envelope, despite the fact that merging BHs with $m_1 \gtrsim 20
\msun$ are mostly formed through stable mass transfer channel
\citep[e.g.][]{2022ApJ...931...17V}. This may come from the fact that
we calculate stellar mass and radii from our single evolution model,
while we obtain convective envelope mass and radii from the original
{\tt BSE} formulae. Nevertheless, even if the bumps are artifacts, the
global maximum of the rate at the primary BH mass of $\sim 9-10$
$\msun$, and the global gradient of the primary BH mass distribution
in the binary BH merger rate are still consistent with GW
observations. Moreover, we place high priority on the reproduction of
the global structure, especially the the PI mass gap BHs. This is
because the PI mass gap BHs are directly related to PISNe, the main
topic of this paper.

We have performed additional calculations by replacing the fiducial
PPISN model with two other PPISN models. The first one is constructed
by \cite{2022RNAAS...6...25R} (hereafter Renzo PPISN model), which is
based on \cite{2019ApJ...887...53F}. The second one is called ``the
moderate PPISN model'' by \cite{2020A&A...636A.104B}, which is based
on the hydrodynamics simulations of \cite{2019ApJ...887...72L}. PPISNe
leave BHs with different masses in the two models, while PPISNe leave
only $45$ $\msun$ BHs in the fiducial models. In this sense, the two
PPISN models are more realistic than the fiducial models. For Renzo
model with $\mmax=150$ $\msun$ and moderate PPISN model with
$\mmax=150$ and $300$ $\msun$, the sharp peak at $45$ $\msun$ in
Figure \ref{fig:mergerRateFromFileMass} is smoothed out. This is a
preferable feature from point of view of comparison with the GW
observations. On the other hand, for Renzo model with $\mmax=300$
$\msun$, the sharp peak survives, and is shifted to $\sim 55$
$\msun$. Although Renzo model is more realistic than the fiducial
model, the primary BH mass distribution deviates from the GW
observations. This means that we cannot increase $\mmax$ freely for
the fiducial models. We have decided to test $\mmax = 150$ and $300$
$\msun$ in order to consider a wide range of possibilities, because
the primary BH mass distribution in the moderate PPISN model is in
good agreement with the GW observations at least.

Our models reproduce merging binary BHs with PI masses even with the
fiducial setup.  Here, we overview the formation path of such binary
BHs. Details of the population synthesis model and related discussions
are found in \cite{2021MNRAS.505.2170T,2022ApJ...926...83T}.  Let us
consider Pop III binary stars consisting of $\sim 90 \msun$ and $\sim
60 \msun$ stars separated by $\sim 90 \rsun$. The heavier star has a
$\sim 40 \msun$ helium core at its post main-sequence phase, while it
expands only to $\sim 40 \rsun$ because of the lack of heavy elements
in the atmosphere and of inefficient convective overshoot. The star
remains compact and experiences little mass transfer. Note that such a
star does not experience common envelope evolution even if it fills
its Roche lobe, because it has a radiative envelope and is stable to
Roche lobe overflow. With the small-mass helium core surrounded by a
massive hydrogen envelope, the star finally collapses to form a $\sim
90 \msun$ BH without triggering PPISN nor PISN.  The lighter
(companion) star also evolves through similar processes, and collapses
to a $\sim 60 \msun$ BH.  During the binary evolution, the separation
does not change significantly because of the absence of mass loss and
mass transfer, and thus the two remnant BHs can merge within a Hubble
time. It is worth noting that this particular formation process does
not occur for binary stars with $\gtrsim 10^{-3} \zsun$
\citep{2022ApJ...926...83T}. A $\sim 90 \msun$ star with $\gtrsim
10^{-3} \zsun$ expands to more than $1000 \rsun$ before collapsing to
a BH, and loses its hydrogen envelope through mass transfer or common
envelope evolution.  In our fiducial models, merging binaries of PI
mass BHs are not produced for $\gtrsim 10^{-3} \zsun$.

From all the current models, we take PISN data including the redshift
evolution of PISN rate density, and properties of PISN progenitors,
such as masses and stellar types. These information is fed into the
\euclid survey simulations as conducted by \cite{2022A&A...666A.157M}.
In short, \euclid plans to observe their Deep Field
($40~\mathrm{deg^2}$ in total) approximately every half year. Each
visit reaches around 25.5~mag in the $I_{\scriptscriptstyle\rm E}$
band and 24.0~mag in the $Y_{\scriptscriptstyle\rm E}$,
$J_{\scriptscriptstyle\rm E}$, and $H_{\scriptscriptstyle\rm E}$
bands. We refer to \cite{2022A&A...666A.157M} and
\cite{2021arXiv210801201S} for the details of the \euclid
observational plan.

By taking the \euclid observational plan presented in
\cite{2022A&A...666A.157M}, we use the aforementioned PISN rate
densities to estimate the expected number of PISN discoveries in each
model. We repeat the survey simulations with the same setup for $10^3$
times and report their average discovery numbers.  In model series
``fid'', once a PISN explode in our survey simulations, we take the
PISN light-curve model of the closest mass from
\cite{2011ApJ...734..102K} and estimate its following brightness
evolution. In model series ``$\tsig$'', we do not have corresponding
PISN explosion models to estimate their luminosity evolution. For
simplicity, we take the closest mass PISN model from
\cite{2011ApJ...734..102K} to estimate the light-curve evolution even
in the case of model series ``$\tsig$''.  The PISN mass range of model
series ``fid'' is $65-135 \msun$ while the PISN mass range for model
series ``$\tsig$'' is $90-180 \msun$. All PISNe above $135 \msun$ in
model series ``$\tsig$'' are approximated by the $130 \msun$ PISN
model from model series ``fid''.  We note that some PISNe above $130
\msun$ may become brighter than $130 \msun$ to be detected at higher
redshifts, but their number density is not high enough to affect our
conclusion in this paper.

\section{Results}
\label{sec:Results}

\begin{figure}
  \includegraphics[width=\columnwidth]{\fdir/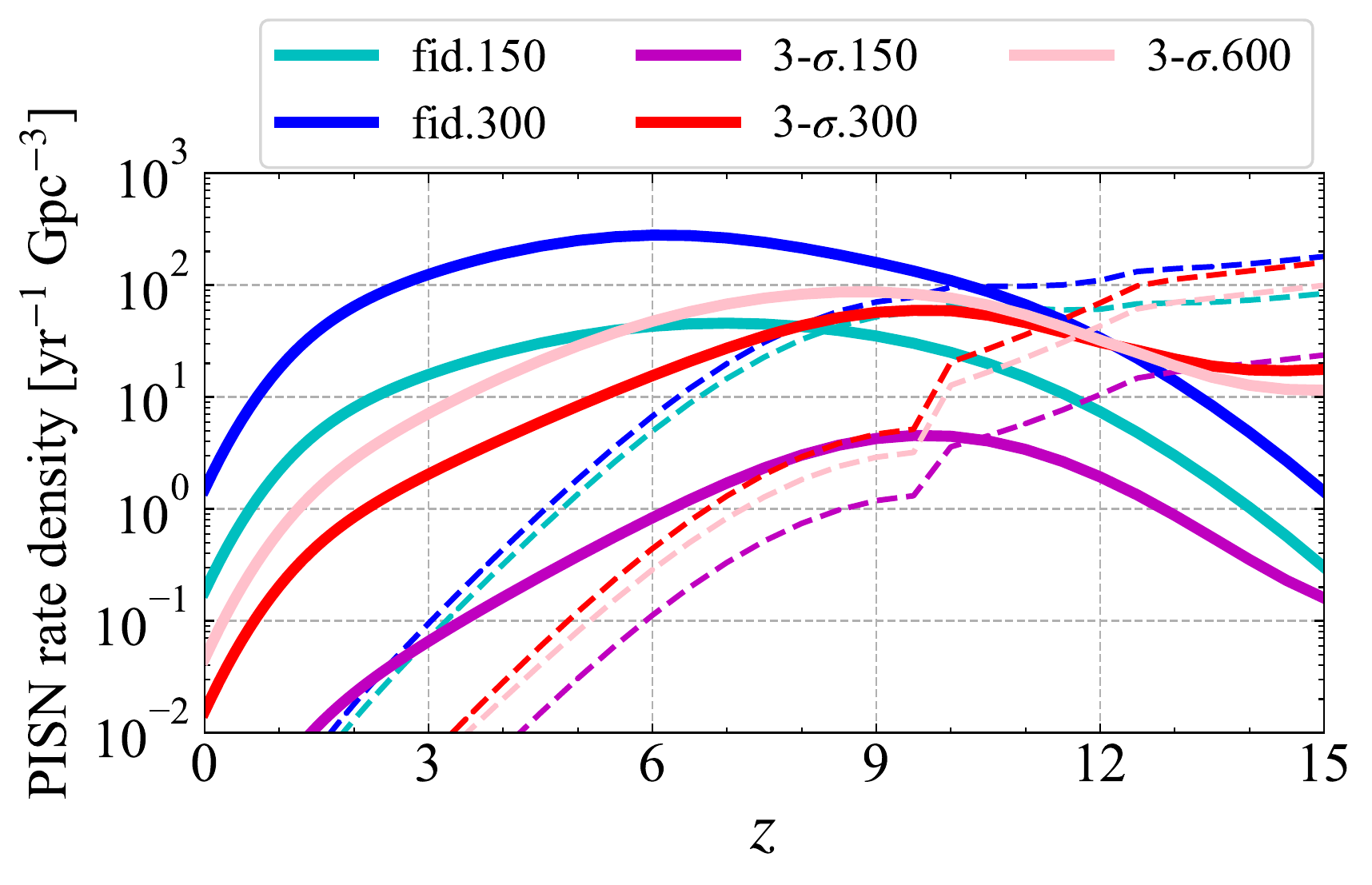}
  \caption{Redshift evolution of PISN rate density for the current 5
    models. Solid and dashed curves indicate Type I and II PISNe,
    respectively, defined in the main text.}
  \label{fig:pisnIntrinsicRateFromFile}
\end{figure}

Figure \ref{fig:pisnIntrinsicRateFromFile} shows the redshift
evolution of PISN rate density for the current 5 models. The solid and
dashed curves indicate Type I and II PISNe, respectively, where Type I
and II PISNe are hydrogen-poor and hydrogen-rich, and have naked
helium stars and post main-sequence stars with hydrogen envelopes as
their progenitors, respectively. For any models, Type I PISNe become
dominant at $z \lesssim 5$. This is because $\gtrsim 99$ \% of stars
have $Z/\zsun \gtrsim 0.1$ at $z \lesssim 5$, and strip off their
hydrogen envelopes due to their stellar winds. The rate densities of
Type II PISNe increase with redshift increasing, and exceed those of
Type I PISNe at $z \gtrsim 10$. At $z \gtrsim 10$, Pop III stars
dominate Type II PISNe. Because of the emergence of Pop III stars, we
can see sudden increase of the rate densities of Type II PISNe at $z
\sim 10$ for model series ``$\tsig$''. There is no such increase for
model series ``fid''. The rate densities of Type II PISNe are high
enough to hide the emergence of Pop III stars in model series
``fid''. Nevertheless, Pop III stars also dominate Type II PISNe at $z
\gtrsim 10$ in model series ``fid''.

We compare our PISN rate density with those in
\cite{2022MNRAS.514.1315B}. Our fid.300 model roughly corresponds to
their ``empirical'' model, in which they adopt the SFH model of
\cite{2014ARA&A..52..415M} that is close to our H22 SFH at $z \lesssim
8$. Their $\mmax$ is $300 \msun$. Our PISN rate density is about 2 or
3 times smaller than theirs according to their figure 9. We find that
this is because our overall metallicity is higher than theirs, and
that our adopted stellar evolution model allows massive stars to lose
more mass through stronger stellar winds. This argument is consistent
with the fact that our binary BH merger rate density is also about 5
times smaller than theirs at $z \lesssim 2$ as seen in their figure
11. If we lowered our metallicity overall, we could obtain larger
binary BH merger rate density than shown in Figure
\ref{fig:mergerRateFromFileTotal}. However, we remark that our binary
BH merger rate density is consistent with that inferred by GWTC-3 with
respect to the global maximum of the rate at the primary BH mass of
$\sim 9-10$ $\msun$, and the global gradient of the primary BH mass
distribution in the binary BH merger rate.

\begin{figure}
  \includegraphics[width=\columnwidth]{\fdir/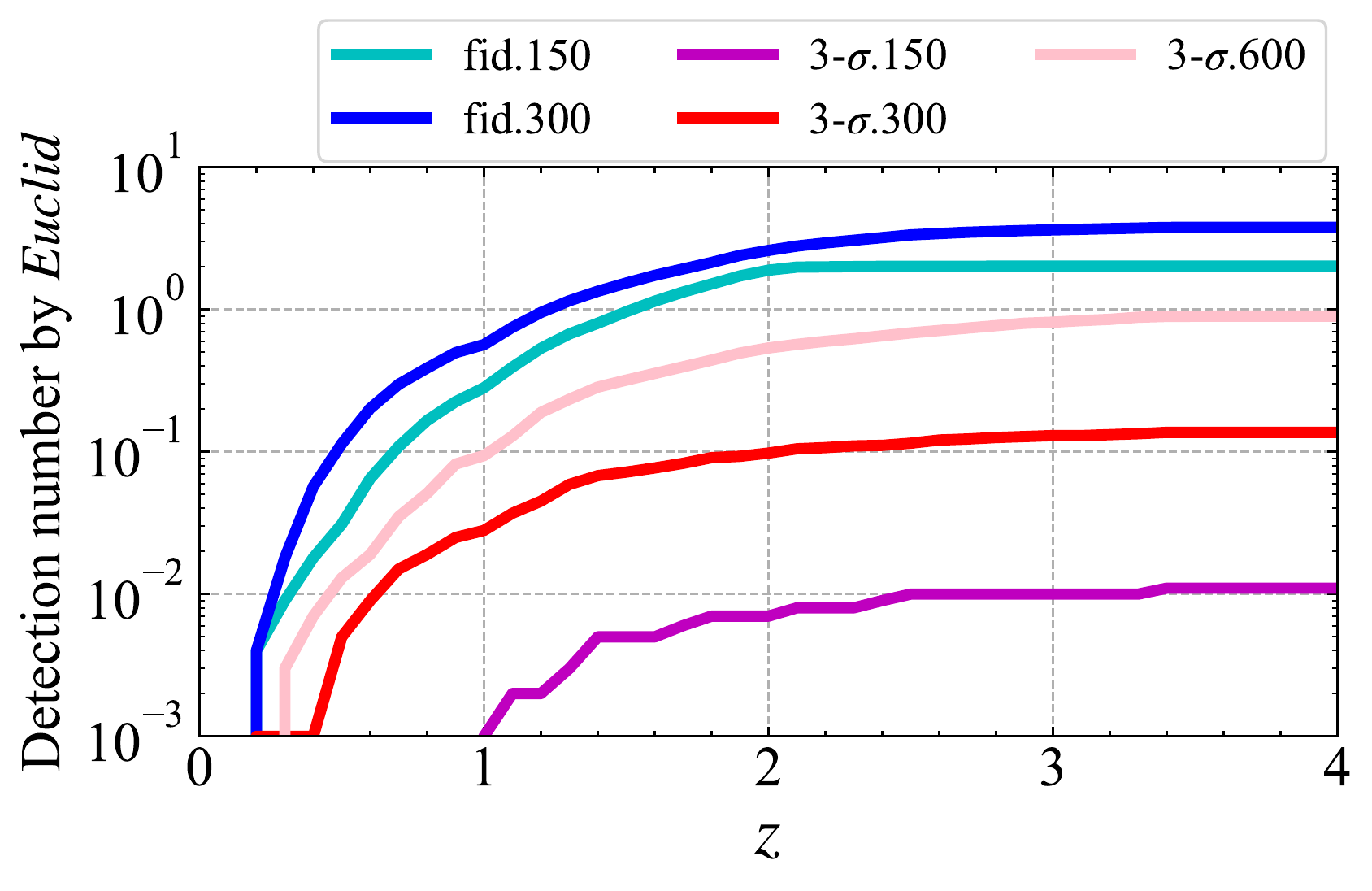}
  \caption{Expected PISN detection numbers to redshift $z$ during
    6-year \euclid operation.}
  \label{fig:pisnDetectionRateFromFile}
\end{figure}

Hereafter, we focus on Type I PISNe, since \euclid will be able to
discover PISNe at such low redshifts. We can see in Figure
\ref{fig:pisnDetectionRateFromFile} the expected PISN detection
numbers within a given $z$ during 6-year \euclid operation. In all the
cases, the detection numbers saturates beyond $z \sim 3$, despite that
the Type I PISN rate densities increase beyond $z \sim 6$. This is
because the PISN detection horizon of \euclid is $z \sim 3$.

The expected detection numbers in model series ``fid'' are not
sensitive to $\mmax$. In model series ``$\tsig$'', the detection
numbers increase with $\mmax$ increasing, but remain smaller than in
model series ``fid'' as can be seen by comparing model $\tsig$.600
with fid.150. This is because PISNe in model series ``$\tsig$'' need
more massive progenitors than in model series ``fid''. In summary, for
$\mmax=150-300 \msun$, the detection numbers in model series ``fid''
and ``$\tsig$'' are much more and less than 1, respectively. On the
other hand, the detection number can be about 1 in model series
``$\tsig$'' only if $\mmax=600 \msun$.

\section{Conclusion and discussion}
\label{sec:ConclusionAndDiscussion}

We exploit binary population synthesis models consistent with binary
BHs observed by GWs with respect to the global maximum of the rate at
the primary BH mass of $\sim 9-10$ $\msun$, and the global gradient of
the primary BH mass distribution in the binary BH merger rate to infer
the expected PISN detection number by \euclid. We find that the
expected PISN detection number is greater than 1 in all our ``fid''
models, but it is significantly smaller in ``$\tsig$'' models under a
reasonable assumption of $\mmax \lesssim 300 \msun$. Thus, when
\euclid discovers several or more PISNe, the "fid" model series is
clearly preferred, and we can also constrain the $\cago$ reaction
rate.

If \euclid discovers just 1 PISN, it may be difficult to identify the
correct model only from the PISN detection number. This is because the
expected PISN detection number is close 1 in our model ``$\tsig$''
with $\mmax \gtrsim 600 \msun$. We cannot rule out the possibility of
$\mmax \sim 600 \msun$, although such massive stars have not been
discovered. According to the present day mass functions in star
clusters R136 and 30 Dor, $\mmax$ appears to be at least $\sim 300
\msun$ \citep{2013A&A...558A.134D, 2018Sci...359...69S,
  2018A&A...618A..73S}. Moreover, the PISN detection numbers of both
the ``fid'' and ``$\tsig$'' models can increase (or decrease) with
stellar winds weakening (or strengthening). Unfortunately, there
remains substantial uncertainty in the strength of stellar winds
\citep[see][for review]{2021ARA&A..59..337D, 2022ARA&A..60..203V}. In
order to solve this degeneracy, PISN ejecta mass will be helpful. For
example, if a future observation discovers a type I PISN with ejecta
mass of $180 \msun$, our "$\tsig$" model series would be preferred.
PISN ejecta mass can be estimated if we can obtain their spectra
around the luminosity peak to constrain their expansion velocity, in
addition to the light curves obtained by the \euclid observations to
break the degeneracy between ejecta mass and explosion energy
\citep[e.g.][]{2008MNRAS.383.1485V}.  Spectroscopic follow-up
observations of PISN candidates are essential.

After we can solve the degeneracy, we can also obtain $\mmax$. As seen
in Figure \ref{fig:pisnDetectionRateFromFile}, the expected PISN
detection number depends on $\mmax$. Note that the wind mass loss rate
of a massive star is highly uncertain
\citep[e.g.][]{2021A&A...647A..13G, 2021MNRAS.504..146V}, and would
affect the detection number. The uncertainty would make the measure of
$\mmax$ difficult, since the detection numbers are similar in the
fiducial models with different $\mmax$ especially for $z \lesssim 2$.

We make caveats about constraints on the formation mechanism of binary
BHs. Even if model series ``fid'' are correct, it does not always mean
that Pop III binary stars form PI mass gap BHs. To be so, Pop III
stars have to evolve with inefficient convective overshoot
\citep{2021MNRAS.505.2170T, 2021ApJ...910...30T,
  2022ApJ...926...83T}. This can be constrained by other PISN surveys
as studied in our future work.

\section*{Acknowledgments}

We are grateful for the anonymous referee for many helpful
suggestions. AT appreciates T. Yoshida and K. Takahashi for many
fruitful advices. We thank organizers of the first star and first
galaxy conference 2021 in Japan for giving us a good opportunity to
start our collaboration. This research could not been accomplished
without the support by Grants-in-Aid for Scientific Research
(17H06360, 19K03907) from the Japan Society for the Promotion of
Science.  This work is supported by JSPS Core-to-Core Program
(JPJSCCA20210003).

\section*{Data availability}

Results will be shared on reasonable request to authors. The results
include the additional calculations by the fiducial PPISN model with
two other PPISN models: the Renzo's PPISN and moderate PPISN models,
which are discussed in section \ref{sec:Method}.

\bsp	
\label{lastpage}
\end{document}